\documentclass[conference]{IEEEtran}
\IEEEoverridecommandlockouts
\usepackage{cite}
\usepackage{amsmath,amssymb,amsfonts}
\usepackage{algorithmic}
\usepackage{graphicx}
\usepackage{textcomp}
\usepackage{xcolor}
\def\BibTeX{{\rm B\kern-.05em{\sc i\kern-.025em b}\kern-.08em
    T\kern-.1667em\lower.7ex\hbox{E}\kern-.125emX}}
\usepackage{caption,subcaption}
\usepackage{amssymb}
\usepackage{xifthen}
\usepackage{siunitx} 
\usepackage{booktabs}
\usepackage{array}
\usepackage{dcolumn}
\usepackage{colortbl}
\makeatletter
\newcommand*{\rom}[1]{\expandafter\@slowromancap\romannumeral #1@}
\makeatother
\usepackage{tikz}
\usepackage{tikz,pgfplots} %drawings and plots
\usepackage{tikz-3dplot} 
\pgfplotsset{compat=newest}
\usepackage[american,siunitx]{circuitikz} %circuit
\usetikzlibrary{decorations}
\usetikzlibrary{decorations.pathreplacing,decorations.markings}
\usetikzlibrary{arrows,positioning}

\usepackage{etoolbox}

\begin{document}

\title{Electromagnetic Sensor and Actuator Attacks on Power Converters for Electric Vehicles}

\author{\IEEEauthorblockN{G\"{o}k\c{c}en Y{\i}lmaz Dayan{\i}kl{\i}*
}
\thanks{*Dayan{\i}kl{\i} and Hatch are co-first authors.}
 \IEEEauthorblockA{
 \textit{Virginia Tech}\\
%  Arlington, VA \\
 gyd@vt.edu}
 \and
 \IEEEauthorblockN{Rees R. Hatch*}
 \IEEEauthorblockA{
 \textit{Utah State University}\\
%  Logan, UT\\
 rees.hatch@usu.edu}
 \and
 \IEEEauthorblockN{Ryan M. Gerdes}
 \IEEEauthorblockA{
 \textit{Virginia Tech}\\
%  Arlington, VA \\
 rgerdes@vt.edu}
 \and
 \IEEEauthorblockN{Hongjie Wang}
 \IEEEauthorblockA{
 \textit{Utah State University}\\
%  Logan, UT \\
 hongjie.wang@usu.edu}
 \and
 \IEEEauthorblockN{Regan Zane}
 \IEEEauthorblockA{
 \textit{Utah State University}\\
%  Logan, UT \\
 regan.zane@usu.edu}
 }

\makeatletter
\patchcmd{\@maketitle}
  {\addvspace{0.5\baselineskip}\egroup}
  {\addvspace{-1.5\baselineskip}\egroup}
  {}
  {}
  
\maketitle

\makeatother
\begin{abstract}
Alleviating range anxiety for electric vehicles (i.e., whether such vehicles can be relied upon to travel long distances in a timely manner) is critical for sustainable transportation.  Extremely fast charging (XFC), whereby electric vehicles (EV) can be quickly recharged in the time frame it takes to refuel an internal combustion engine, has been proposed to alleviate this concern.  A critical component of these chargers is the efficient and proper operation of power converters that convert AC to DC power and otherwise regulate power delivery to vehicles.  These converters rely on the integrity of sensor and actuation signals. In this work the operation of state-of-the art XFC converters is assessed in adversarial conditions, specifically against Intentional Electromagnetic Interference Attacks (IEMI). The targeted system is analyzed with the goal of determining possible weak points for IEMI, viz.\ voltage and current sensor outputs and gate control signals. This work demonstrates that, with relatively low power levels, an adversary is able to manipulate the voltage and current sensor outputs necessary to ensure the proper operation of the converters. Furthermore, in the first attack of its kind, it is shown that the gate signal that controls the converter switches can be manipulated, to catastrophic effect; i.e., it is possible for an attacker to control the switching state of individual transistors to cause irreparable damage to the converter and associated systems. Finally, a discussion of countermeasures for hardware designers to mitigate IEMI-based attacks is provided.

% significant voltage (\%100) and current (\%30) sensor output manipulations can be achieved with relatively low power levels (\SI{200}{\milli\watt} and ).

% and current sensor outputs can be manipulated by \%100  and by with a power level of \SI{200}{\milli\watt}. On the other side, during the current sensor output attack, it is observed that PCB traces can be a target for IEMI attacks. In the final scenario, the attacker injects voltage to control 

% system description is made to determin is made and design Power converter systems Intentional Elelctromagnetic Interference (IEMI) attacks targeting sensor and actuators of power converters is a significant threat for secure operation of these devices. We analyzed the resilience of a  

% for electrical vehicles are significant threat

% Realization of Sensor and Actuator Attacks on Power Converters for Electric Vehicles

% Electromagnetic Sensor and Actuator Attacks on Power Converters for Electric Vehicles

% Vulnerability Assessment of Power Converters for Electric Vehicles

% Intentional Electromagnetic Interference (IEMI) Attacks on Power Converter for Electric Vehicle Extreme Fast Charger
\end{abstract}

\section{Introduction}
In order to increase the adoption of electric vehicles (EV) it is necessary that extremely fast chargers (XFC), along with the attendant battery management systems (BMS), be developed.  These advances in charging technology will ensure that EVs can be charged in a time frame commensurate with that of refilling an internal combustion engine vehicle, and therefore alleviate concerns vehicle owners have regarding the feasibility of using EV for routine and long distance travel.  The security of XFC chargers and BMS are of great importance since attacks on these systems could result in the overcharging of the EV battery (leading to, e.g., potential fire). Larger scale synchronized attacks on XFC chargers, since they connect the EV to the power grid, could cause instability in the grid leading to blackouts.

Until now the security of EV power converter systems has been largely ignored. In this work we seek to enhance the security of EVs by examining the potential vulnerabilities of XFC chargers and BMS.  To this end we provide simulation and experimental results for attacks against critical components of the systems, namely their sensor and actuator (switching) capabilities. For the first time, we demonstrate electromagnetic-based, non-intrusive attacks on actual power converters (comprising AC-DC and BMS power converters) and discuss possible countermeasures. 

\subsection{Related Work}
IEMI is known to be an important threat for analog sensor readings in the security literature. IEMI attacks have been reported on light sensors, temperature sensors, speed sensors, implantable cardiac devices and microphones \cite{Selvaraj2018,Tu2019,Shoukry2013,Kune2013}. Although each attack starts with injecting radiation at the resonance frequency of the targeted device, device-specific non-linearities, due to amplifiers \cite{Tu2019,Kune2013} and ADCs \cite{Selvaraj2018}, can be exploited by attackers to manipulate the sensor data. The reader is referred to \cite{giechaskiel2019sok} for a comprehensive review of such attacks. Since amplifiers and ADCs are commonly used in power converters for sensing and feedback control, IEMI can be used to attack XFC power converters with both relatively low-cost and low-power.  

\subsection{Contributions}
 In this work we examine the vulnerability of state-of-the art XFC power converter designs to IEMI attacks. To the best of authors' knowledge, this is the first study that focuses on power converter security from the perspective of IEMI attacks. Specifically, we
 demonstrate three attacks to show that both the sensing and actuator signals of power converters can be manipulated via non-invasive means (i.e., no physical connection with the hardware are necessary, thereby allowing for proximate attacks).  Our primary contributions are: 
 \begin{itemize}
     \item Showing that the voltage and current sensor outputs of power converters, necessary to maintain the proper and safe control of the converters, can be manipulated with low-cost and low-power amplifiers and radiators. 
     \item Demonstrating that, and proving an analytical model that explains how, drivers/switches can be controlled (i.e., open or closed) via difficult to shield IEMI. Such drivers/switches are ubiquitous in hardware and cyber-physical systems and we are the first to show and explain how their proximate manipulation may be effected. 
     \item Proposing several widely applicable design changes to hardware level to mitigate IEMI attacks.
\end{itemize}
Attacks are experimentally validated and, for safety's sake, their affects demonstrated in simulation via Matlab Simulink.
\section{System Models}

\label{sec:System}
The targeted/victim system consists of an extreme fast charger (XFC) and battery management system (BMS). The XFC is a high-power (\SI{350}{\kilo\watt}) converter designed to convert 3-phase AC power into DC voltage for EV charging; thus, it is known as an AC to DC (AC-DC) converter.  As the power-level increases, the battery charging process poses potential safety risks to EV users in the event that an adversary gains control of the system, as described later.  This section describes the AC-DC and BMS, their controls, and weak points from a theoretical perspective. 

\begin{figure}[!t]
\centering
\includegraphics[width=8.8cm]{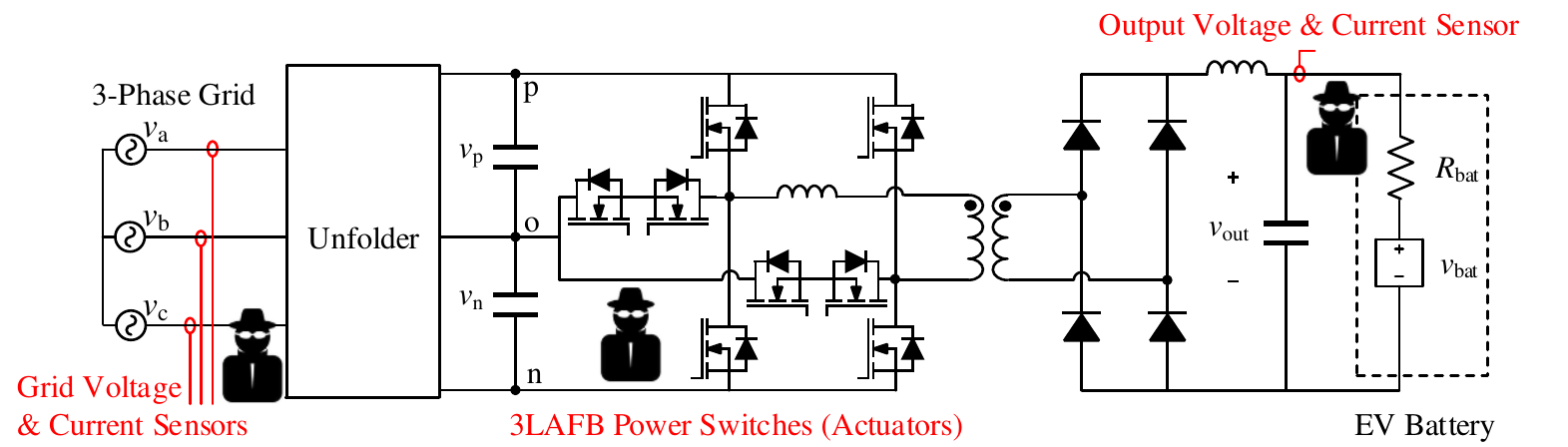}
\caption{Functional diagram of the 3-phase AC-DC converter with 3LAFB topology.  The figure highlights attack points in red, viz.\ the output and grid voltage and current sensors as well as the gate signals to the power switches (actuators).}
\label{fig:Functional}
\end{figure}

The specific AC-DC converter analyzed in this paper is the 3-Level Asymmetric Full Bridge (3LAFB) \cite{Yelaverthi2019}, which is an isolated converter topology intended for use in Unfolding based rectifiers.  The  functional diagram of the Unfolder and 3LAFB topology is shown in Fig.\ \ref{fig:Functional}.  Based upon a safety analysis, the diagram identifies the most sensitive points of attack to be the voltage and current sensors used to monitor the converter inputs and outputs, as well as the power switches.  
The control objective of the AC-DC converter is to regulate the charging of EV batteries.  Battery charging is typically implemented in a constant current constant voltage (CC-CV) scheme. The EV battery is charged at a constant current until the max battery voltage is reached. The charger then switches to constant voltage (CV) control until the battery is fully charged.  It is important to note that EV batteries subjected to charging currents or voltages greater than allowable values cause the cells to overheat which creates a fire hazard. 

The control of the AC-DC is achieved by switching the 3LAFB to regulate average voltage and current.  The feedback sensors are commonly implemented by low voltage analog hardware that is digitized by an ADC. The controller updates the duty cycle for switches based on the sensed error. (The duty cycle determines the average amount of time the switches are turned on in one switching period.)  In actuality, individual power transistors are turned on and off by gate drivers driven by pulse width modulation (PWM) signals. 

The switches and their gate drivers can be thought of as the system actuators because they actuate the PWM gate signals from a micro-controller. The gating signals, being PWM signals, command the transistor to turn on (logic high) or off (logic low).  The 3LAFB has 8 transistors and 8 gating signals while the unfolder requires 12 of each. Gate drivers operate similar to transistors in that they require the input signal to rise above a certain threshold voltage in order to change the devices switching state.

The system's weak points, with respect to IEMI, lie within the feedback sensors and low-voltage gating signals. The converter can only regulate the output correctly if the feedback voltage/current sensors are measuring accurately.  Furthermore, the system can only be controlled if the correct gate signal from the controller is being acted upon by the switches.  Thus, large enough disruptions in the gating signal (\SI{3.3}{\volt} logic) can cause the gate driver to actuate a false turn-on or turn-off of a power switch.

The BMS operates on the same principles as the AC-DC converter.  The purpose of the BMS, comprised of multiple DC-DC converters, is to balance the individual cells that make up an EV battery-pack. Each DC-DC converter has its own voltage and current sensors that measure the flow of power for that cell.  The BMS employs a current and/or voltage feedback loop for each DC-DC by controlling the duty cycle (or equivalent control signal).  
\section{Attack Simulations and Outcomes}
\label{sec:Outcome}
To explore the effects of IEMI attacks on the battery charging operation of the AC-DC, the attack scenario is simulated in Matlab.  The system is modeled on a switching level using PLEC's Blockset add-on for Simulink.  The hardware parameters from a \SI{2}{\kilo\watt} prototype \cite{Yelaverthi2019} were used for the simulation.  The operating point for the simulation is given in Table \ref{tab60}.  The 3LAFB attack is implemented at a DC operating point where the input voltages of the 3LAFB are held constant at a particular grid phase angle rather than the time-varying input that occurs during normal AC operation.  The AC input should be considered when the attackers target the grid voltage and current sensors which will affect the Unfolder operation and AC-DC power quality. Due to space constraints, only the CV regulator will be investigated; however, the presented analysis can be extended to other parts of the system.
\begin{table}[b]
\caption{Operating Point for CV IEMI Simulations}
\begin{center}
\begin{tabular}{|c|c|c|c|}
\hline
\textbf{Parameter}&\textbf{Value}&\textbf{Parameter}&\textbf{Value} \\
\hline
$V_{bat}$ & \SI{500}{\volt} & $R_{bat}$ & \SI{0.5}{\ohm} \\
\hline
$V_{out,ref}$ & \SI{502}{\volt} & $\phi_{grid}$ & \ang{45}  \\
\hline
$V_{p}$ & \SI{480}{\volt} & $V_{n}$ & \SI{176}{\volt} \\
\hline
\end{tabular}
\label{tab60}
\end{center}
\end{table}

\begin{figure}[!t]
\centering
\includegraphics[width=8.8cm]{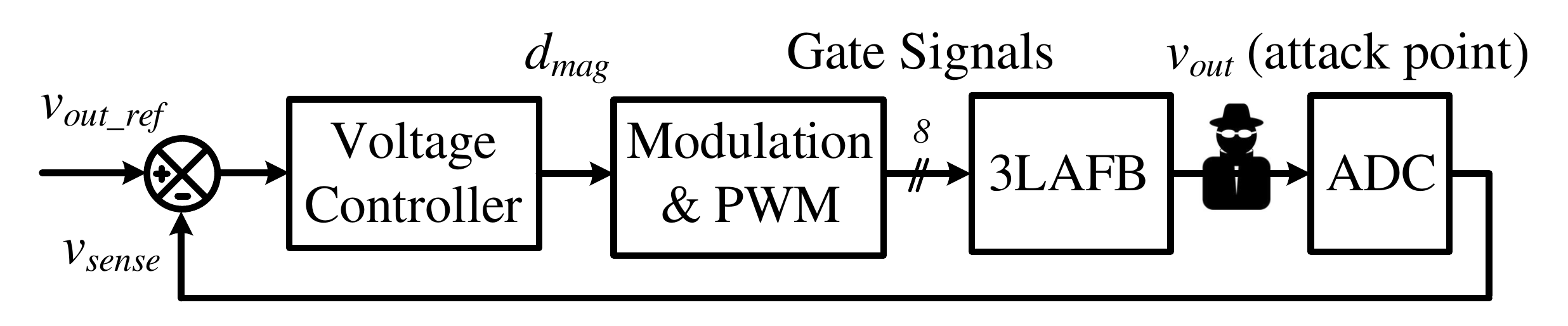}
\caption{Block diagram of the constant voltage controller for the 3LAFB. The attacker targets the analog circuitry before the sensing information is digitized by the ADC.}
\label{fig:control_diagram}
\end{figure}

Based on an efficiency and safety analysis of the system, we consider a scenario wherein an attacker is able to overcharge the battery by manipulation of the power converter's feedback voltage signal. Such over-voltage charging would lead to increased charging current at the maximum voltage.  The extra power dissipated as heat by the resistive losses of the battery would cause cell heating.  Repeated attacks of this nature would lead to decreased battery capacity and lifespan.  In the extreme case, where the battery is subjected to sustained over-current charging, the increase in cell temperatures could lead to thermal runaway in which the battery pack would ignite and create a self-sustaining fire. To cause damage to the battery it is simply necessary to subvert CV control (specifically the second phase of the CC-CV charging scheme). %The battery current in this mode is only limited by the battery resistance and can be calculated as:
%
%\begin{equation}
%i_{bat}(t) = \frac{v_{out}(t) - v_{bat}(t)}{R_{bat}}
%\label{eqn:Ibat}
%\end{equation}
%
The CV control loop demonstrated in Fig.\ \ref{fig:control_diagram} uses feedback from the output voltage sensor to control the magnitude of applied duty cycles, $d_{mag}$.  In this scenario, the attacker is targeting the $v_{sense}$ which is the sensed feedback signal of $v_{out}$, the output voltage of the converter.

\begin{figure}[b]
\centering
\resizebox{8.8cm}{!}{%
\definecolor{mycolor1}{rgb}{0.00000,0.44700,0.74100}%
\definecolor{mycolor2}{rgb}{0.85000,0.32500,0.09800}%

\begin{tikzpicture}

%\draw [decorate,decoration={brace,amplitude=10pt,mirror,raise=4pt}]
\draw [decorate,decoration={brace,mirror,amplitude=4pt}]
  (1.25,1) -- (1.25,1.25) node[black,midway,xshift=0.75cm] {\footnotesize $\SI{17.9}{\milli\volt}$};

\begin{axis}[%
width=3.022in,
height=0.581in,
at={(0.344in,1.102in)},
scale only axis,
xmin=0,
xmax=60,
ymin=501.287214029935,
ymax=503.076539462723,
ylabel style={font=\bfseries\color{white!15!black}},
ylabel={Volts},
ylabel absolute,
axis background/.style={fill=white},
axis x line*=bottom,
axis y line*=left,
xmajorgrids,
ymajorgrids,
grid style={dotted},
]
\addplot [color=mycolor1, line width=2.0pt]
  table[row sep=crcr]{%
0	502.086942476317\\
9.89999999999998	502.087817167154\\
10.05	502.287214029935\\
10.2	502.468473076994\\
10.35	502.505026666072\\
10.5	502.52752087476\\
10.8	502.566529633695\\
11.1	502.601738216418\\
11.55	502.650395404182\\
12	502.694471304008\\
12.45	502.733847512832\\
12.9	502.769148610051\\
13.35	502.80084988659\\
13.8	502.829280006738\\
14.25	502.854779809089\\
14.85	502.884746479529\\
15.45	502.910671398316\\
16.05	502.933100249608\\
16.65	502.95250450406\\
17.4	502.973122486041\\
18.15	502.990325655595\\
19.05	503.007251489418\\
19.95	503.020871761725\\
21	503.033438236626\\
22.2	503.044393110222\\
23.7	503.05429744644\\
25.5	503.062296728705\\
27.75	503.068456410099\\
30.75	503.072855865168\\
35.25	503.075597570042\\
39.9	503.076539462723\\
40.05	502.880121354671\\
40.2	502.698443264448\\
40.35	502.661839399101\\
40.5	502.639226606887\\
40.8	502.6000473245\\
41.1	502.564943404587\\
41.55	502.516539539065\\
42	502.472537618163\\
42.45	502.433192720408\\
42.9	502.397915778944\\
43.35	502.366204903248\\
43.8	502.337742436956\\
44.25	502.31219612453\\
44.85	502.282147514123\\
45.45	502.256125043914\\
46.05	502.233588933475\\
46.65	502.214071936139\\
47.4	502.19331054837\\
48.15	502.175965628005\\
49.05	502.158876566745\\
49.95	502.145103912931\\
51	502.132375926132\\
52.2	502.121259190772\\
53.7	502.111185731948\\
55.5	502.103027342754\\
57.75	502.096723882836\\
60	502.093048255715\\
};
% \addlegendentry{data1}

\addplot [color=mycolor2, dashed, line width=2.0pt]
  table[row sep=crcr]{%
0	502.086942476317\\
9.89999999999998	502.087817167154\\
10.05	501.287214029935\\
10.2	501.468473076994\\
10.35	501.505026666072\\
10.5	501.52752087476\\
10.8	501.566529633695\\
11.1	501.601738216418\\
11.55	501.650395404182\\
12	501.694471304008\\
12.45	501.733847512832\\
12.9	501.769148610051\\
13.35	501.80084988659\\
13.8	501.829280006738\\
14.25	501.854779809089\\
14.85	501.884746479529\\
15.45	501.910671398316\\
16.05	501.933100249608\\
16.65	501.95250450406\\
17.4	501.973122486041\\
18.15	501.990325655595\\
19.05	502.007251489418\\
19.95	502.020871761725\\
21	502.033438236626\\
22.2	502.044393110222\\
23.7	502.05429744644\\
25.5	502.062296728705\\
27.75	502.068456410099\\
30.75	502.072855865168\\
35.25	502.075597570042\\
39.9	502.076539462723\\
40.05	502.880121354671\\
40.2	502.698443264448\\
40.35	502.661839399101\\
40.5	502.639226606887\\
40.8	502.6000473245\\
41.1	502.564943404587\\
41.55	502.516539539065\\
42	502.472537618163\\
42.45	502.433192720408\\
42.9	502.397915778944\\
43.35	502.366204903248\\
43.8	502.337742436956\\
44.25	502.31219612453\\
44.85	502.282147514123\\
45.45	502.256125043914\\
46.05	502.233588933475\\
46.65	502.214071936139\\
47.4	502.19331054837\\
48.15	502.175965628005\\
49.05	502.158876566745\\
49.95	502.145103912931\\
51	502.132375926132\\
52.2	502.121259190772\\
53.7	502.111185731948\\
55.5	502.103027342754\\
57.75	502.096723882836\\
60	502.093048255715\\
};
% \addlegendentry{data2}

\addplot [color=mycolor1, dashed, line width=2.0pt]
  table[row sep=crcr]{%
0	502\\
60	502\\
};
% \addlegendentry{data3}

\end{axis}

\begin{axis}[%
width=3.022in,
height=0.581in,
at={(0.344in,0.295in)},
scale only axis,
xmin=0,
xmax=60,
xlabel style={font=\bfseries\color{white!15!black}},
xlabel={Time [ms]},
ymin=4,
ymax=6,
ylabel style={font=\bfseries\color{white!15!black}},
ylabel={Amps},
ylabel absolute,
axis background/.style={fill=white},
xmajorgrids,
ymajorgrids,
grid style={dotted},
]
\addplot [color=mycolor1, line width=2.0pt]
  table[row sep=crcr]{%
0	4.00760777307113\\
7.8	4.00923556374347\\
9.90000000000001	4.00935292575593\\
10.05	4.40358801374555\\
10.2	4.77092570889884\\
10.35	4.84463997015695\\
10.5	4.88959014849262\\
10.65	4.92969533337209\\
10.95	5.00339959563392\\
11.25	5.07174987088969\\
11.55	5.13616141880737\\
11.85	5.19639716667307\\
12.15	5.25228341348722\\
12.45	5.30414751906489\\
12.75	5.35241799873773\\
13.05	5.39737759314142\\
13.35	5.43921878537476\\
13.65	5.47813701634706\\
14.1	5.53148599339488\\
14.55	5.57936596244199\\
15	5.62233208811686\\
15.45	5.66088576204123\\
15.9	5.69548022598813\\
16.35	5.72652111551471\\
16.8	5.7543725276235\\
17.4	5.78710709766823\\
18	5.81543395045889\\
18.6	5.83994584303419\\
19.35	5.86599600624595\\
20.1	5.88773575138201\\
20.85	5.90587770183969\\
21.75	5.92372971192233\\
22.8	5.94020267389809\\
24	5.95456479697341\\
25.35	5.96645359907691\\
26.85	5.97582865204804\\
28.8	5.98389573781967\\
31.2	5.98979686020952\\
34.65	5.99404601914055\\
39.9	5.99639444847311\\
40.05	5.60172762546016\\
40.2	5.23266101202475\\
40.35	5.15860144118705\\
40.5	5.11330003974362\\
40.65	5.07294069311516\\
40.95	4.99926860937347\\
41.25	4.93136788265473\\
41.55	4.86735987874943\\
41.85	4.80739227503406\\
42.15	4.75174966151003\\
42.45	4.70016163893542\\
42.75	4.65217743651743\\
43.05	4.6074861822531\\
43.35	4.56589234358323\\
43.65	4.52720748754998\\
44.1	4.47418517851185\\
44.55	4.42659764916199\\
45	4.38388906680776\\
45.45	4.34556058614657\\
45.9	4.31116057461501\\
46.35	4.28028549858488\\
46.8	4.25257373088188\\
47.4	4.21998905833206\\
48	4.19177592582745\\
48.6	4.16734713734105\\
49.35	4.14136538184487\\
50.1	4.11966283501463\\
50.85	4.10153418924344\\
51.75	4.08367522051061\\
52.8	4.06717287291512\\
54	4.05276114977684\\
55.35	4.04080795479054\\
56.85	4.03136093076263\\
58.8	4.02320955120671\\
60	4.01979026098709\\
};
% \addlegendentry{data1}

\end{axis}

\draw [decorate,decoration={brace,mirror,amplitude=4pt}]
  (4.5,3.5) -- (4.5,4.2) node[black,midway,xshift=0.75cm] {\footnotesize $\SI{1}{\volt}$ attack};
  
\draw [decorate,decoration={brace,mirror,amplitude=4pt}]
(4,0.9) -- (4,2.1) node[black,midway,xshift=1cm] {\footnotesize $\SI{2}{\ampere}$ response};
  
\draw[solid]
  (6.2,1.9) -- (6.5,1.9);
\node at (7,1.9) {$i_{out}$};

\draw[solid]
  (2.3,3.7) -- (2.7,3.7);
\node at (3.2,3.7) {$v_{out}$};

\draw[solid]
  (2.7,3.1) -- (3,3.1);
\node at (3.5,3.1) {$v_{sense}$};

\draw[solid]
  (6.2,3.1) -- (6,3.3);
\node at (6.8,3.1) {$v_{out\_ref}$};

\end{tikzpicture}%
}
\caption{Simulation of constant voltage controller attack.  An attacker induces a \SI{1}{\volt} offset into the $v_{sense}$ signal causing the charging current to increase from 4 to \SI{6}{\ampere}, which indicates that a small change in sensed voltage can lead to a substantial increase in current (and thus heating of a battery).}
\label{fig:CV_attack}
\end{figure}
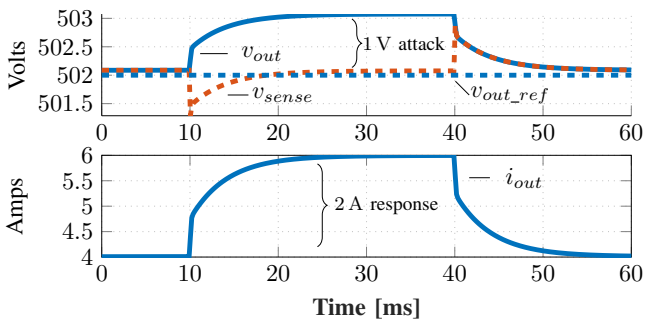

In the simulation shown in Fig.\ \ref{fig:CV_attack} an IEMI attack is initiated on the hardware at \SI{10}{\milli\second}.  The attack is simulated by altering the feedback signal, $v_{sense}$, by subtracting \SI{1}{\volt} from the actual output voltage; i.e., the attacker decreases the apparent output voltage which will cause the control system to compensate by increasing the output voltage. This alteration represents the average voltage distortion that is induced on an ADC sensor used to measure output voltage during an IEMI attack.  The simulated attack is sustained for \SI{30}{\milli\second}. 

As can be seen from the figure, the controller regulates the sensed voltage to the reference voltage of \SI{502}{\volt}; however, the actual output voltage is \SI{503}{\volt}.  On the short time scale of the simulation, the battery voltage is approximately constant and \SI{500}{\volt}. The extra \SI{1}{\volt} on the output causes the battery current to increase from 4 to \SI{6}{\ampere}, a significant increase in current that would cause heating.  The charging current is extremely sensitive to changes in $v_{out}$ due to the small battery resistance ($<$\SI{1}{\ohm}), which implies that small changes in sensed voltage result in geometrical increases in current (and thus heat).  
\section{Theory of Attack}
\label{sec:Theory}
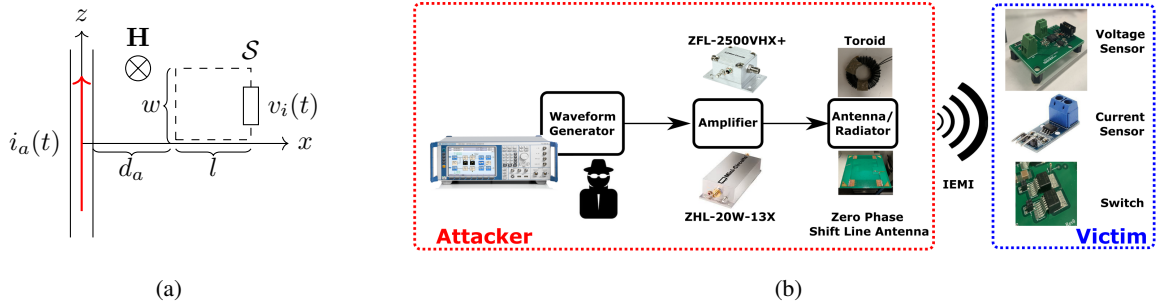
\begin{figure*}[!t]
\centering
\subcaptionbox{\label{fig:EMModel}}{\raisebox{0.1\height}{\begin{tikzpicture}[cross/.style={path picture={\draw[black] (path picture bounding box.south east) -- (path picture bounding box.north west) (path picture bounding box.south west) -- (path picture bounding box.north east);}}]

%z,x
\draw[->] (0,0) -- (0,1.5);
\node[anchor=south] at (0,1.5) {$z$};

\draw[->] (0,0) -- (2.75,0);
\node[anchor=west] at (2.75,0) {$x$};

%current
\draw (-0.15,-1.25) -- node[left] {$i_{a}(t)$} (-0.15,1.25);
\draw[->,red,thick] (0,-0.9) -- (0,0.9);
\draw (0.15,-1.25) -- (0.15,1.25);

%surface
\draw[dashed] (1.25,0.05) -- (1.25,1) -- (2.25,1) -- (2.25,0.05) --
(1.25,0.05);
\node[anchor=south] at (2.25,1) {$\mathcal{S}$};

%load
\filldraw[fill=white, draw=black] (2.15,0.25) rectangle
(2.35,0.75);
\node[anchor=west] at (2.35,0.5) {$v_{i}(t)$};

%field
\node [draw,circle,cross,minimum width=0.25 cm,label={$\mathbf{H}$}](B) at (0.75,1){}; 

%dimensions
\draw[decorate,decoration={brace,mirror,raise=2pt},black] (0.15,0) --
node[below,black,yshift=-2pt] {$d_{a}$} (1.18,0);
\draw[decorate,decoration={brace,mirror,raise=2pt},black] (1.25,0) --
node[below,black,yshift=-2pt] {$l$} (2.25,0);
\draw[decorate,decoration={brace,raise=2pt},black] (1.25,0) --
node[left,black,xshift=-2pt] {$w$} (1.25,1);

\end{tikzpicture}}}
\hspace{0.1\columnwidth}%
\subcaptionbox{\label{fig:AttackerHardware}}{\includegraphics[width=0.55\linewidth]{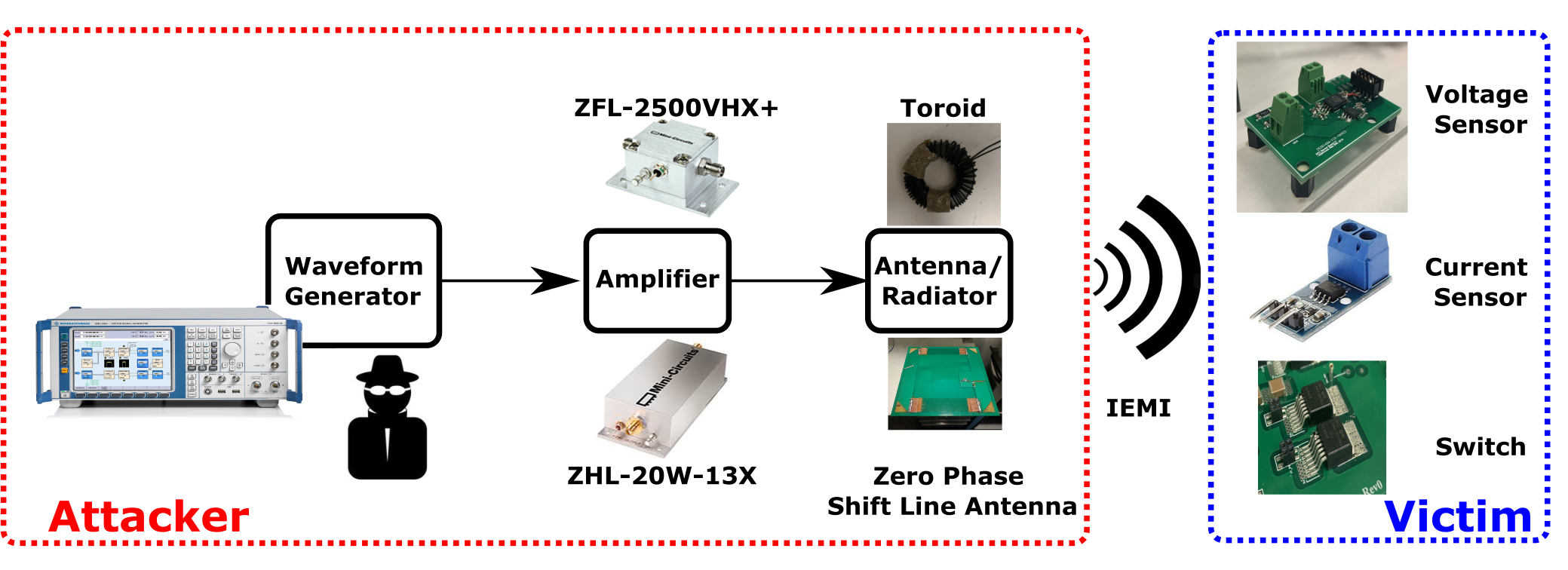}}
%\hspace{-5pt}
\caption{(a) IEMI attack model \cite{Selvaraj2018} (b) Attacker hardware and attack points for power converters}
\end{figure*}

Our attacks are based on Faraday's law of induction, which states that a time varying magnetic field captured by a conducting loop results in a voltage on the loop \cite{Maxwell2010}.  By such means are we able to modify the voltages measured by sensors, and used to control switches, in power converters. To observe how a time varying current, $i_{a}$, supplied by an attacker, induces a voltage, $v_{i}$, on a victim loop, an infinitely long, z-axis directed current is assumed to be positioned at distance $d_a$ from the victim circuit having dimensions $w$ and $l$ (Fig.\ \ref{fig:EMModel}). By Faraday's law and Ampere's law the relationship between the attacker signal, $i_{a}$, and the induced voltage,  $v_{i}$, is: 
\begin{equation}
v_{i}(t)=-\mu \left[\frac{w}{2\pi}\ln\left(\frac{d_{a}+l}{d_{a}}\right)\right] \frac{d}{dt}i_{a}(t)
\label{eqn:vinduced}
\end{equation}
where the permeability of the medium is $\mu$.

% \begin{equation}
% v_{induced}(t)=\oint E \cdot dl =-\int_S \frac{\partial \mathbf{B}} {\partial t} \cdot d\mathbf{S}\\
% \label{eqn:Faraday}
% \end{equation}

% where $E$, $B$ and $v_{induced}$ are the electric field, magnetic field and induced voltage on the victim loop, respectively.

% \begin{figure}[!thb]
% \centering
% \subcaptionbox{\label{fig:EMModel}}{\raisebox{\dimexpr.5\ht1-.5\height}{\input{Figure/EMModel}}}
% \caption{Model for analytical solution \cite{Selvaraj2018}.}
% \end{figure}

% \begin{figure}[!b]
% \centering
% \input{Figure/EMModel}
% \caption{Model for analytical solution \cite{Selvaraj2018}.}
% \end{figure}

% \begin{figure*}[!thb]
% \centering
% \includegraphics[width=0.9\textwidth]{Figure/AttackerHardware}
% \caption{The attacker hardware and attack points}
% \label{fig:AttackerHardware}
% \end{figure*}

The amplitude and shape (waveform) of $v_{i}$ are determined by a geometry coefficient (square brackets) and the time derivative of $i_{a}$. In the following attack scenarios, the attacker uses a continuous sinusoidal $i_{a}$ attack waveform, so the form of $v_{i}$ is a sinusoidal with a phase shift due to transmitting hardware. We note that an increased victim loop size results in an increase induced $v_{i}$.

\subsection{Threat Model} 
We assume an attacker aiming to manipulate the operation of an AC-DC converter and BMS through IEMI. It is assumed that the attacker can place EM radiators in proximity to the converters but there is no physical connection between the attacker hardware and victim circuitry. The attacker has access to commodity RF component and devices, e.g., waveform generators, RF amplifiers and EM radiators like toroids and antennas (Figure \ref{fig:AttackerHardware}).  We consider an attacker who targets weak points of the victim system using a toroid with a focused magnetic field or a ZPSL antenna with a directive near field radiation pattern. The weak points discussed in detail in Section \ref{sec:System} are chosen as attack points (voltage sensor output,$v_{out}$, BMS current sensor output,$i_{cell}$, and the low voltage gate signals that control the AC-DC switches).

\subsubsection{Attack Point \rom{1} - Voltage Sensor Output}
The attacker uses IEMI to manipulate the voltage sensor data $v_{out}$ by inducing voltage $v_i$ on the victim cable that connects the analog sensor output and the ADC input of the CV controller. The attack has two phases: the first phase is the efficient EM coupling to the victim cable through the use of cable resonant frequency as an attack frequency \cite{Kune2013}. Before each attack, a frequency sweep is applied to detect the resonant frequency of the victim cable. The next phase is the manipulation of non-linearity of ADC. An ADC samples and digitizes an analog signal in the ADC input range ($v_{min}$ to $v_{max}$). A very common practice is to average the digitized data to filter out high frequency noise. It is discussed in \cite{Selvaraj2018} how a generic ADC transfer function and electrostatic discharge (ESD) diodes result in a phenomenon called clipping. We assume the input voltage of the ADC is compromised and a time varying voltage $v_{ADC}$ is fed into the ADC as follows:
\begin{equation}
v_{ADC}(t)=V_{s}+v_{i}(t)
\label{eqn:Vadc}
\end{equation}
where $V_{s}$ is a relatively low frequency sensor output which is assumed as a DC offset and $v_i$ is a purely sinusoidal induced signal by IEMI with frequency $f$ and amplitude $V_i$. For small sensor output $V_s$ close to $v_{min}$ case, we assumed that $v_{i}=\SI{0}{\volt}$. In that case, the measured voltage by the ADC has the form of a half wave rectified signal assuming $V_i<v_{max}$. The average value (DC) of a half wave rectified sinusoidal waveform with amplitude $V_i$ and period $T=1/f$ is:
\begin{equation}
V_{DC}=\frac{1}{T}\left(\int_{0}^{\frac{T}{2}}V_i sin(2\pi f t) \text{d}t + \int_{\frac{T}{2}}^{T}0 \text{d}t\right)= \frac{V_i}{\pi}
\label{eqn:HalfWaveDC}
\end{equation}
Note that Equation \ref{eqn:HalfWaveDC} assumes an infinite sampling frequency and ignores the effects which is observed when the attack frequency is a perfect multiple of sampling frequency (i.e., relative phase becomes important). Other affects also render Equation \ref{eqn:HalfWaveDC} an approximation that works well in practice; the reader is referred to \cite{Selvaraj2018} for a detailed discussion of inducing DC voltages via AC signals. %discusses the effects of states that when the analog sensor output is \SI{0}{\volt} or equal to $v_{min}$, the average ADC manipulation is $V_i/\pi$. However, this increasing effect is observed less intensely with increasing $V_S$ up to $v_{max}/2$. For $V_S$ values larger than $v_{max}/2$, the effect of sinusoidal waveform $v_i$ is to decrease the ADC outputs, which can be explained by clipping of positive side of the induced $v_i$ \cite{Selvaraj2018}. 

\subsubsection{Attack Point \rom{2} - Current Sensor Output}
This attack point consists of the PCB trace between the analog current sensor output and the input of controller ADC (Figure \ref{fig:control_diagram}). It is assumed that the attacker can place the EM radiator (e.g., an air gap toroid) to induce a high magnetic field. The two phase attack mechanism that includes the efficient coupling and manipulation of the ADC discussed in the previous section is applicable in this attack as well. However, this attack has a fundamental difference: the attack point is a PCB trace which requires the manipulation of smaller victim loops than \textit{Attack \rom{1}} and necessitates higher attack powers. 

\subsubsection{Attack Point \rom{3}-Gate Control Signal}
\begin{figure}[!b]
\centering
\includegraphics[width=0.4\textwidth]{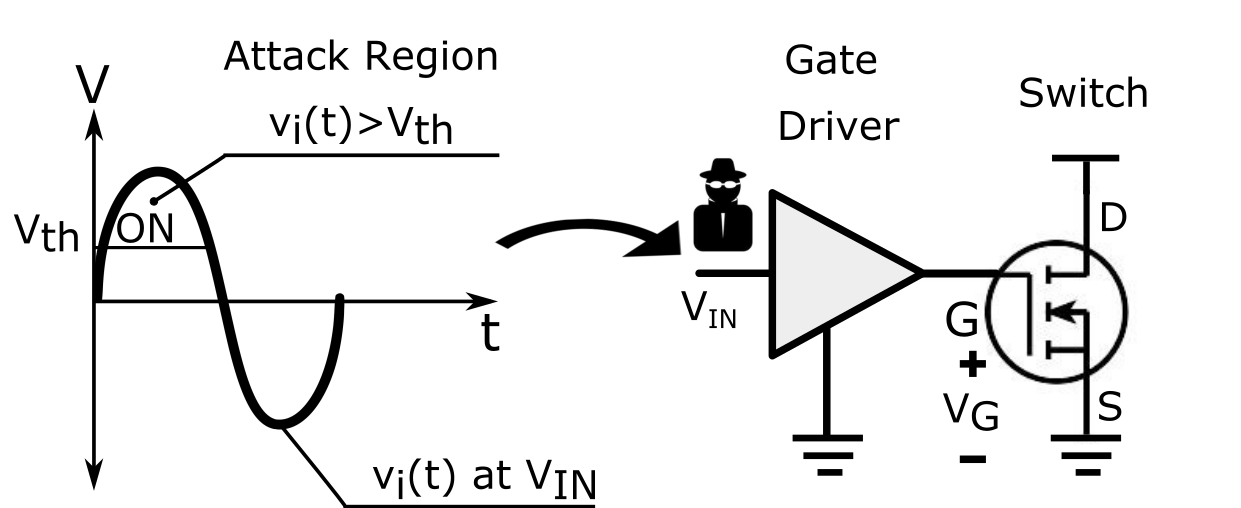}
\caption{The induced voltage $v_i(t)$ to $V_{IN}$ should exceed the gate driver threshold voltage $V_{th}$ to turn the switch on.}
\label{fig:SwitchMec}
\end{figure}

The 3LAFB employs a high current gate driver \cite{GateDriver} that controls an SiC switch \cite{Transistor} as shown in Figure \ref{fig:SwitchMec}. The attacker aims to change the input voltage $V_{IN}$ of gate driver to control the switch. To turn on the gate driver and switch, the attacker should satisfy the condition in \ref{eqn:TurnOn} which is also demonstrated in Figure \ref{fig:SwitchMec}:
\begin{equation}
v_i(t) = V_i sin(2\pi f t) > V_{th}\,\,\, \text{Switch ON}
\label{eqn:TurnOn}
\end{equation}
 where $v_i$ is the voltage induced at the input of the gate driver and $V_{th}$ is the minimum voltage to activate the gate.

\section{Experimental Results}
\label{sec:Experimental}
Three attack points are experimentally tested against IEMI.

\subsection{Attack \rom{1}: False Voltage Sensor Data Injection}
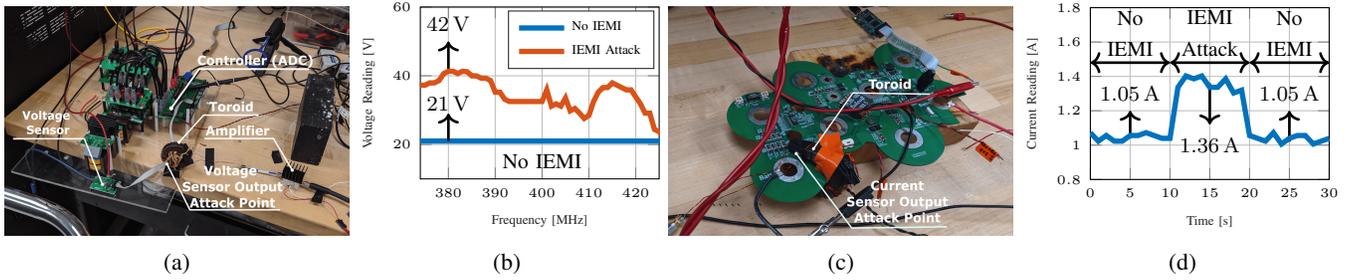
\begin{figure*}[!t]
\centering
\subcaptionbox{\label{subfig:VoltageSetup}}{\includegraphics[height=1.2in, width=1.8in]{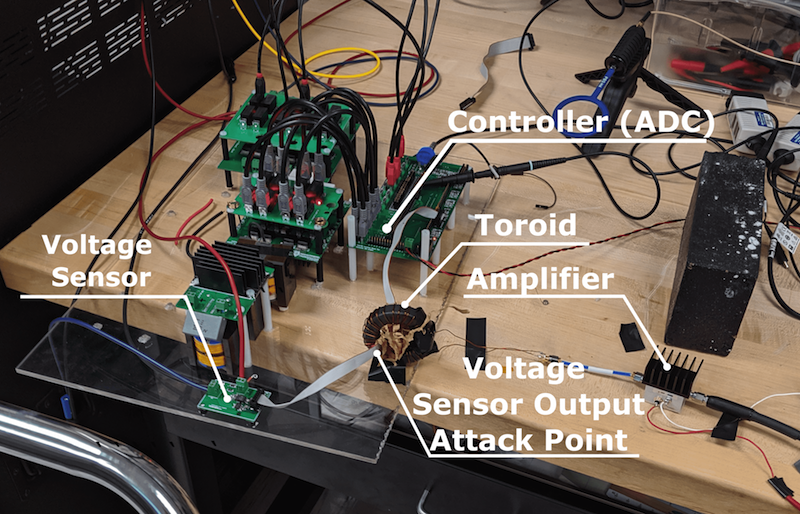}}
\hspace{-5pt}
\subcaptionbox{\label{subfig:VoltageReadings}}{\raisebox{-1\height}{% This file was created by matlab2tikz.
\definecolor{mycolor1}{rgb}{0.00000,0.44700,0.74100}%
\definecolor{mycolor2}{rgb}{0.85000,0.32500,0.09800}%
\begin{tikzpicture}
\pgfplotsset{every tick label/.append style={font=\tiny}}
\begin{axis}[%
width=1.25in,
height=0.9in,
at={(0.5in,0in)},
scale only axis,
xmin=374,
xmax=425,
xtick = {380,390,400,410,420},
xlabel style={font=\tiny\color{white!15!black}},
xlabel={Frequency [MHz]},
ymin=10,
ymax=60,
ylabel style={font=\tiny\color{white!15!black}},
ylabel={Voltage Reading [V]},
axis background/.style={fill=white},
xmajorgrids,
xminorgrids,
ymajorgrids,
legend style={font=\tiny,legend cell align=left, align=left, draw=white!15!black}
]
\addplot [color=mycolor1, line width=2.0pt]
  table[row sep=crcr]{%
374	21\\
375	21\\
376	21\\
377	21\\
378	21\\
379	21\\
380	21\\
381	21\\
382	21\\
383	21\\
384	21\\
385	21\\
386	21\\
387	21\\
388	21\\
389	21\\
390	21\\
391	21\\
392	21\\
393	21\\
394	21\\
395	21\\
396	21\\
397	21\\
398	21\\
399	21\\
400	21\\
401	21\\
402	21\\
403	21\\
404	21\\
405	21\\
406	21\\
407	21\\
408	21\\
409	21\\
410	21\\
411	21\\
412	21\\
413	21\\
414	21\\
415	21\\
416	21\\
417	21\\
418	21\\
419	21\\
420	21\\
421	21\\
422	21\\
423	21\\
424	21\\
425	21\\
426	21\\
427	21\\
428	21\\
429	21\\
};
\addlegendentry{No IEMI}

\draw [->,line width=1pt] (380,42) -- (380,50);
\node[align=center] at (380,55) {\footnotesize \SI{42}{\volt}};

\draw [->,line width=1pt] (380,21) -- (380,29);
\node[align=center] at (380,32) {\footnotesize \SI{21}{\volt}};
\node[align=center] at (400,15) {\footnotesize No IEMI};

\addplot [color=mycolor2, line width=2.0pt]
  table[row sep=crcr]{%
374	37.258064516129\\
375	37.258064516129\\
376	37.9354838709677\\
377	37.9354838709677\\
378	38.6129032258064\\
379	40.6451612903226\\
380	41.3225806451613\\
381	41.3225806451613\\
382	40.6451612903226\\
383	41.3225806451613\\
384	41.1870967741935\\
385	39.9677419354839\\
386	39.9677419354839\\
387	39.2903225806452\\
388	39.2903225806452\\
389	38.6129032258064\\
390	35.2258064516129\\
391	35.2258064516129\\
392	33.1935483870968\\
393	32.5161290322581\\
394	32.5161290322581\\
395	32.5161290322581\\
396	32.5161290322581\\
397	32.5161290322581\\
398	32.5161290322581\\
399	32.5161290322581\\
400	32.5161290322581\\
401	35.9032258064516\\
402	31.1612903225806\\
403	31.8387096774194\\
404	31.1612903225806\\
405	33.5322580645161\\
406	29.8064516129032\\
407	30.4838709677419\\
408	29.1290322580645\\
409	27.5709677419355\\
410	28.4516129032258\\
411	32.5161290322581\\
412	35.2258064516129\\
413	36.5806451612903\\
414	36.5806451612903\\
415	37.9354838709677\\
416	37.258064516129\\
417	36.5806451612903\\
418	35.9032258064516\\
419	35.9032258064516\\
420	33.8709677419355\\
421	31.8387096774194\\
422	29.1290322580645\\
423	29.1290322580645\\
424	24.3870967741935\\
425	23.7096774193548\\
426	22.3548387096774\\
427	21\\
428	21\\
429	21\\
};
\addlegendentry{IEMI Attack}

\end{axis}
\end{tikzpicture}%
}}
\subcaptionbox{\label{subfig:CurrentSetup}}{\includegraphics[height=1.2in, width=1.8in]{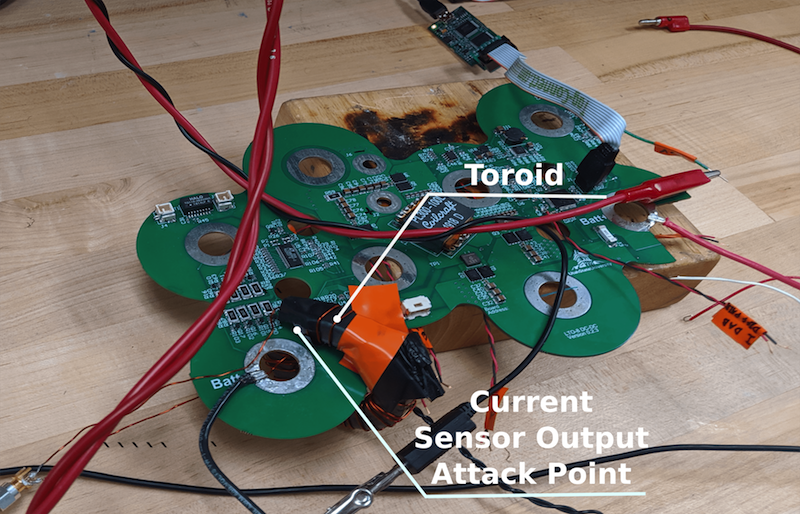}}
\hspace{-5pt}
\subcaptionbox{\label{subfig:CurrentReadings}}{\raisebox{-1\height}{% This file was created by matlab2tikz.
\definecolor{mycolor1}{rgb}{0.00000,0.44700,0.74100}%
\begin{tikzpicture}
\pgfplotsset{every tick label/.append style={font=\tiny}}
\begin{axis}[%
width=1.25in,
height=0.9in,
at={(0.758in,0in)},
scale only axis,
xmin=0,
xmax=30,
xtick = {0,5,10,15,20,25,30},
xlabel style={font=\tiny\color{white!15!black}},
xlabel={Time [s]},
ymin=0.8,
ymax=1.8,
ylabel style={font=\tiny\color{white!15!black}},
ylabel={Current Reading [A]},
axis background/.style={fill=white},
xmajorgrids,
ymajorgrids
]

%IEMI Text
\node[align=center] at (5,1.65) {\footnotesize No\\\footnotesize  IEMI};
\node[align=center] at (15,1.65) {\footnotesize IEMI\\\footnotesize Attack};
\node[align=center] at (25,1.65) {\footnotesize No\\\footnotesize  IEMI};

%Draw arrows
  \draw [<->,line width=1pt] (0,1.48) -- (10,1.48);
  \draw [<->,line width=1pt] (10,1.48) -- (20,1.48);
  \draw [<->,line width=1pt] (20,1.48) -- (30,1.48);
  \draw [->,line width=1pt] (5,1.04) -- (5,1.2);
  \draw [->,line width=1pt] (15,1.36) -- (15,1.12);
  \draw [->,line width=1pt] (25,1.02) -- (25,1.2);
%Current Text
\node[align=center] at (5,1.3) {\footnotesize \SI{1.05}{\ampere}};
\node[align=center] at (15,1) {\footnotesize \SI{1.36}{\ampere}};
\node[align=center] at (25,1.3) {\footnotesize \SI{1.05}{\ampere}};

\addplot [color=mycolor1, line width=2.0pt, forget plot]
  table[row sep=crcr]{%
0	1.0725\\
1	1.023\\
2	1.056\\
3	1.023\\
4	1.023\\
5	1.056\\
6	1.056\\
7	1.0725\\
8	1.056\\
9	1.0395\\
10	1.0395\\
11	1.3365\\
12	1.4025\\
13	1.386\\
14	1.4025\\
15	1.3365\\
16	1.3365\\
17	1.386\\
18	1.3035\\
19	1.32\\
20	1.056\\
21	1.0395\\
22	1.023\\
23	1.0725\\
24	1.0065\\
25	1.0395\\
26	1.056\\
27	1.056\\
28	1.0065\\
29	1.023\\
30	1.0395\\
};
\end{axis}
%Draw arrows
\end{tikzpicture}%
}}
\hspace{-5pt}
\caption{False voltage and current sensor data injection attacks (a) Experimental setup for voltage sensor output manipulation  (b)Voltage sensor output manipulation with regard to attack frequency: measured voltage increased by \SI{21}{\volt} under IEMI (c) Experimental setup for current sensor output manipulation (d) Current readings with regard to time, when IEMI is applied between $t=\SI{10}{\second}$ and  $t=\SI{20}{\second}$, the average of current readings increased from \SI{1.05}{\ampere} to \SI{1.36}{\ampere}} \label{fig:Analog}
\end{figure*}

 The attacker locates the toroid around the victim cable as in Figure \ref{subfig:VoltageSetup}. The toroid has an air-gap which can be filled with a ferrite piece which eliminates the need for the attacker to unplug any wire in the victim. The attacker system consists of a Mini-Circuits ZFL-2500VHX+ RF amplifier and a 30 coil toroid (Figure \ref{fig:AttackerHardware}). The attack power is fixed at \SI{200}{\milli\watt} throughout Attack \rom{1}.

\textbf{Measurement Methodology:} The voltage output of a DC supply is adjusted to \SI{21}{\volt} and connected to the voltage sensor as reference voltage. The system is observed to function properly before the IEMI applied. To magnify the effect of IEMI attack (i.e. less power same data manipulation or same power more data manipulation), an attacker can use the resonant frequency of the victim system as attack frequency \cite{Kune2013}. At resonance the imaginary component of the impedance is minimum, which results in higher induced voltages. To detect the resonant frequency of the victim cable, a frequency sweep between \SI{100}{\mega\hertz} and \SI{500}{\mega\hertz} is applied with \SI{10}{\mega\hertz} increments and voltage sensor data manipulation is observed from a PC. Although all tested attack frequencies result in varying increases in the voltage readings, it is observed that between \SI{380}{\mega\hertz} and  \SI{420}{\mega\hertz}, the effect is more pronounced.

% \begin{figure}[!b]
%     \centering
%     \begin{subfigure}[t]{0.45\textwidth}
%         \centering
%         \includegraphics[width=\textwidth]{Figure/VoltageSetup}
%         \caption{Attack Setup}
%                         \label{subfig:VoltageSetup}
%     \end{subfigure}%
    
%     \begin{subfigure}[t]{0.45\textwidth}
%         \centering
%         \includegraphics[width=\textwidth]{Figure/VoltageIEMI}
%         \caption{Attack frequency versus voltage reading manipulation}
%                 \label{subfig:VoltageReadings}
%     \end{subfigure}
%     \caption{False voltage sensor data injection}
%     \label{fig:VoltageAttack}
% \end{figure}

\textbf{Results:} Figure \ref{subfig:VoltageReadings} shows the voltage reading manipulation under IEMI. Depending on the frequency, the voltage readings are manipulated up to the range between \SI{28}{\volt} and \SI{42}{V}, while the reference voltage is \SI{21}{\volt}. Specifically, at \SI{380}{\mega\hertz}, the voltage reading is increased by \% 100 to \SI{42}{\volt}. Another observation is that the IEMI injection results in an increase of voltage readings throughout the frequency range. This observation is parallel to the ADC nonlinearity discussion in Section \ref{sec:Theory}, as the \SI{21}{\volt} test voltage results in sensor voltages on the lower half of ADC input range. The IEMI on voltage sensor output is a significant threat for a converter because of the low power nature of the attack. On the other side, Simulink analyze shows that even a \SI{1}{\volt} data manipulation can increase the output current significantly (Figure \ref{fig:CV_attack}).
% \vspace*{-0.55\baselineskip}

\subsection{Attack \rom{2}: False Current Sensor Data Injection}
\vspace*{\baselineskip}
The attacker aims to manipulate the current sensor data on the printed circuit board (PCB) of the BMS. The air gapped toroid is positioned on the PCB trace as shown in Figure \ref{subfig:CurrentSetup}. The attacker hardware consists of a \SI{20}{\watt} RF amplifier (Mini-Circuits ZHL-20W-13X) and the toroid. The amplifier output power is adjusted to \SI{2.5}{\watt} to eliminate any mismatch problem due to dominantly imaginary impedance of the toroid. 

% \begin{figure}[!b]
%     \centering
%     \begin{subfigure}[t]{0.45\textwidth}
%         \centering
%         \includegraphics[width=\textwidth]{Figure/CurrentSetup}
%         \caption{Attack Setup}
%         \label{subfig:CurrentSetup}
%     \end{subfigure}
    
%     \begin{subfigure}[t]{0.45\textwidth}
%         \centering
%         \includegraphics[width=\textwidth]{Figure/CurrentIEMI}
%         \caption{Current reading manipulation between $t=\SI{10}{\second}$ and $t=\SI{20}{\second}$}
%         \label{subfig:CurrentReadings}
%     \end{subfigure}
%     \caption{False current sensor data injection }
%     \label{fig:CurrentAttack}
% \end{figure}

\textbf{Measurement Methodology:} The current sensor is supplied with a \SI{1}{\ampere} test current and the system is tested before IEMI radiation. It is observed that the system is operating properly and correct current data is received by the controller.  Then, a sinusoidal EMI with varying frequency between \SI{10}{\mega\hertz} and \SI{500}{\mega\hertz} with \SI{10}{\mega\hertz} increments is applied and it is observed that in the vicinity of \SI{100}{\mega\hertz}, the current data manipulation is much more pronounced. 

\textbf{Results:} In Figure \ref{subfig:CurrentReadings}, the current sensor outputs of the system is provided under a temporary IEMI attack between $t=\SI{10}{\second}$ and $t=\SI{20}{\second}$. The attack frequency is \SI{100}{\mega\hertz}. It is observed that when IEMI starts at $t=\SI{10}{\second}$, the mean value of current readings increase by \% 30 from \SI{1.05}{\ampere} to \SI{1.36}{\ampere}. Note that the test current of \SI{1}{\ampere} is still applied during the attack. On the other side, it is observed that the attack results in an increase in the sensor data which is parallel with the discussion made in Section \ref{sec:Theory}. This attack shows that the PCB traces can be direct targets for IEMI which means PCB level countermeasures are necessary for secure systems.

\subsection{Attack \rom{3}: False Gate Voltage Injection: Turning on Switches with IEMI}

The attacker hardware includes a \SI{20}{\watt} RF amplifier (Mini-Circuits ZHL-20W-13X) and a Zero-Phase-Shift Loop (ZPSL) antenna (Figure \ref{fig:AttackerHardware}). ZPSL antenna is a near field resonant antenna with a strong magnetic field at \SI{72}{\mega\hertz} directed through z axis. The attacker positions the ZPSL antenna \SI{10}{\centi\meter} above intertwined and shielded cables that carry $V_{IN}$ and ground of the gate  driver. We will use the terminology where $V_{IN}$ is the gate driver input or voltage and $V_G$ is switch gate voltage (Figure \ref{fig:SwitchMec}).

\textbf{Measurement Methodology:} 
Attack frequency is chosen as \SI{72}{\mega\hertz} and the attack power is increased by \SI{1}{\decibel} increments from \SI{100}{\milli\watt} to \SI{20}{\watt}, $V_{IN}$ and $V_{G}$ are observed with an oscilloscope. $V_{IN}$ is set to low throughout the measurements which results $V_{G}$ is held at \SI{-3}{\volt} to ensure the switch stays off. If the attack is successful (i.e., switch is turned on by gate drive), the gate voltage $V_G$ is expected to increase to \SI{18}{\volt} by the gate driver. To capture the  turn on characteristic for $V_G$ and $V_{IN}$, the oscilloscope is set to single trigger for a low to high transition at $V_G$.

\begin{figure}[!b]
    \centering
    \begin{subfigure}[t]{0.35\textwidth}
        \centering
        \includegraphics[width=\textwidth]{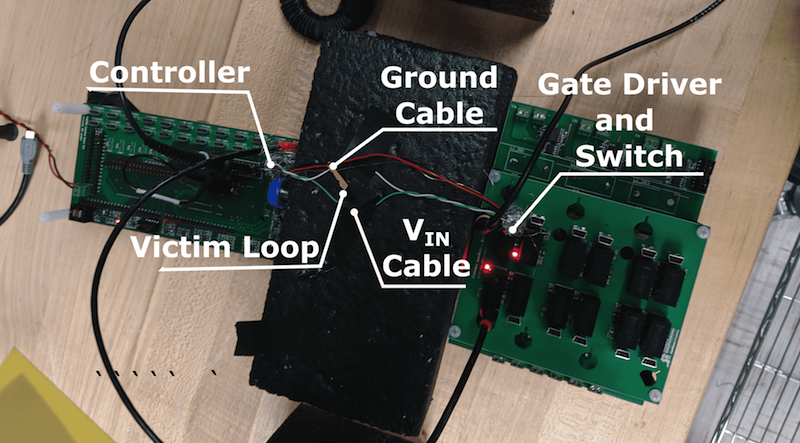}
        \caption{Attack Setup (Antenna is not shown.)}
        \label{subfig:SwitchSetup}
    \end{subfigure}
    
    \begin{subfigure}[t]{0.35\textwidth}
        \centering
        \includegraphics[width=\textwidth]{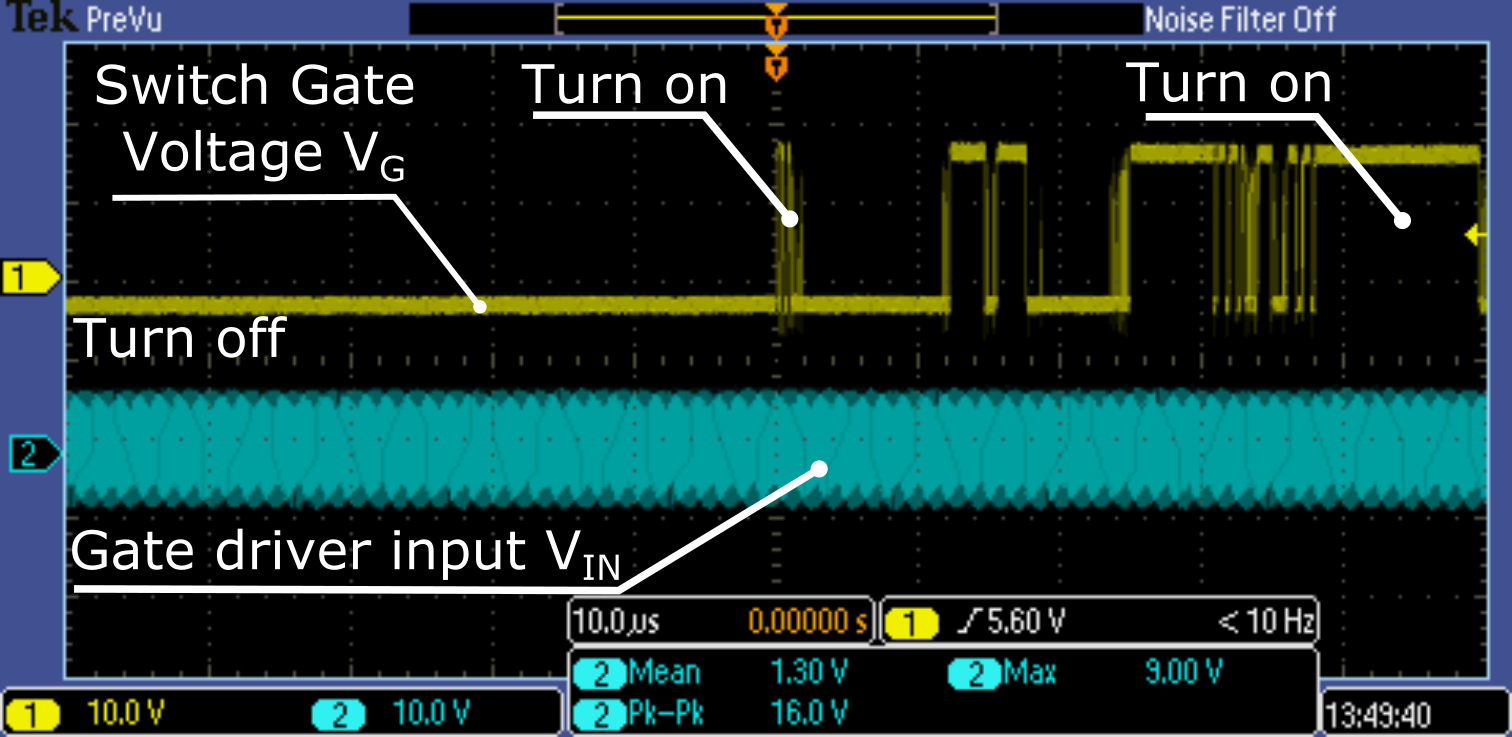}
        \caption{Turn on increment of the transistor}
        \label{subfig:SwitchResult}
    \end{subfigure}
    \caption{False $V_{IN}$ injection: turning on switches with IEMI}
    \label{fig:SwitchAttack}
\end{figure}

\textbf{Results:} When the \SI{20}{\watt} IEMI applied from an attack distance of \SI{10}{\centi\meter}, it is observed that the IEMI is not sufficent to turn on the switch. This is an expected result, because the loop area between cables that carry ground and $V_{IN}$ connection is small and differential voltage between $V_{IN}$ and ground is not high enough to satisfy the condition in Equation \ref{eqn:TurnOn}. Although this shows that sending $V_{IN}$ and ground cables through intertwined cables are relatively secure, in PCB based systems, the $V_{IN}$ and ground traces/pads is not always close due to the minimum spacing requirements of manufacturing process. To observe this phenomenon, the green $V_{IN}$ and the white ground cables are physically separated and a loop of \SI{4}{\centi\metre\squared} is exposed as demonstrated in Figure \ref{subfig:SwitchSetup}. When the attack power is  to \SI{20}{\watt}, it is observed that the $V_G$ increases and switch turns on as shown in yellow plot of Figure \ref{subfig:SwitchResult}. First of all, it is observed that the switch turns on and off until it stabilizes at turn on condition. As we trigger the oscilloscope for a time window of \SI{100}{\micro\second}, the power increase is not observable in the $V_{IN}$ (blue). A possible reason for this phenomenon is the output power increase is smaller than \SI{1}{decibel} as the amplifier operates in saturation.

\section{Discussion of Attacks}
IEMI attacks on the prototype (Section \ref{sec:Experimental}) have exposed potentially catastrophic weaknesses in the AC-DC  and BMS systems.  The ability for the attackers to significantly alter the average ADC values of the power converter's feedback sensors poses a serious threat to the safety of XFC. The $v_{out}$ voltage sensor with a range of \SI{600}{\volt} had an induced error equivalent to \SI{21}{\volt} of error. As was shown in Section \ref{sec:Outcome}, an error of \SI{1}{\volt} in the output voltage sensing was enough to significantly disrupt the operation of the CV controller. 

Every voltage and current sensor used for control in the converter design is a potential weakness to be mitigated. The attacker's ability to control the switches through alteration of the gate signal is another attack point.  The digital gate signals are not as sensitive to the IEMI as the sensed, analog signals; however as was shown, if the victim loop of the gate signal is large enough, the attackers are able to turn on switches that were intended to be closed. If this event occurs on live hardware, a short-circuit event is likely to occur. The incredible currents and heat generated in a short-circuit is likely to cause system wide device failure or at least system shutdown.

\subsection*{Countermeasures}
Although RF shielding (e.g., conductive sheet or foam) is effectively used against relatively high frequency signals, the low frequency ($<\!\!100$ MHz) and magnetic nature of the reported attack signal makes it very difficult to shield fast chargers \cite{Ott2011}. Adding to that, none of the magnetic field shielding options (e.g., MuMetal and Faraday cage) are employed in commercial fast chargers. In order to protect PCB traces transmitting sensitive signals (e.g., analog sensor outputs and gate/switch control signals), hardware designers should be aware of IEMI threats from the first moment of layout generation and eliminate large loops between significant traces and ground pad/traces. However, due to minimum spacing restrictions of PCB manufacturing process and complex layout designs with many components, eliminating large loops may not always possible. In those situations, we suggest using via-fenced striplines for analog sensor outputs and gate driver signals. Although via-fenced stripline is used for eliminating crosstalk between traces, it can also be used to eliminate high frequency IEMI from outside sources. We are also investigating alternative approaches that seek to randomize multiple sections of the pathway signals take from sensor to ADC, controller or actuator that would make the resonant frequency of traces unknown to the attacker and thus limit their ability to couple to circuits and affect signals.
\section{Conclusion}
The AC-DC and Battery Management System (BMS) of the power converter is observed to be vulnerable to IEMI attacks. Both systems rely on feedback of the converter outputs to properly regulate the flow of power in the circuit. The system's low voltage current and voltage sensor outputs and gate control signals are susceptible to IEMI attacks which distort the converter's control by inducing a DC offset to the sensed value. The attackers can gain control of the system by manipulation of the feedback signal and can cause damage to the EV, XFC, and BMS systems with one or combination of attacks. Furthermore, the control signals from the micro-controller to the gate drivers can also be vulnerable given the victim loop and attacker power level is large enough to induce sufficient voltage. As a future work, we plan to investigate additional PCB level countermeasures and produce prototypes to test these ideas. Our end goal is to provide a design guideline for secure PCB layout design against IEMI.

\section*{Acknowledgment}
This work was supported in part by the Department of Energy under grant No.\ DE-EE0008453.

\bibliographystyle{ieeetr}
\bibliography{Paper}

% \bibitem{b1} G. Eason, B. Noble, and I. N. Sneddon, ``On certain integrals of Lipschitz-Hankel type involving products of Bessel functions,'' Phil. Trans. Roy. Soc. London, vol. A247, pp. 529--551, April 1955.
% \bibitem{b2} J. Clerk Maxwell, A Treatise on Electricity and Magnetism, 3rd ed., vol. 2. Oxford: Clarendon, 1892, pp.68--73.
% \bibitem{b3} I. S. Jacobs and C. P. Bean, ``Fine particles, thin films and exchange anisotropy,'' in Magnetism, vol. III, G. T. Rado and H. Suhl, Eds. New York: Academic, 1963, pp. 271--350.
% \bibitem{b4} K. Elissa, ``Title of paper if known,'' unpublished.
% \bibitem{b5} R. Nicole, ``Title of paper with only first word capitalized,'' J. Name Stand. Abbrev., in press.
% \bibitem{b6} Y. Yorozu, M. Hirano, K. Oka, and Y. Tagawa, ``Electron spectroscopy studies on magneto-optical media and plastic substrate interface,'' IEEE Transl. J. Magn. Japan, vol. 2, pp. 740--741, August 1987 [Digests 9th Annual Conf. Magnetics Japan, p. 301, 1982].
% \bibitem{b7} M. Young, The Technical Writer's Handbook. Mill Valley, CA: University Science, 1989.
% \end{thebibliography}

\end{document}